# Interference-Induced Complex Nonlinearities in Metal-ITO Metasurfaces

Christopher E. Stevens[1,2], Matthew Klein[1,2], Ashley Luo[2], Gregory Vatrano[3], Dennis E. Walker, Jr.,[2], Shivashankar R. Vangala[2], Joshua R. Hendrickson[2], Ivan Avrutsky[4,5], Maxim Sukharev[6,7]

[1]KBR, Beavercreek, OH, USA
[2]Air Force Research Laboratory, Sensors Directorate, Wright-Patterson AFB, OH, USA
[3]Department of Physics, Astronomy, and Materials Science, Missouri State University, Springfield, MO 65897, USA
[4]Department of Electrical and Computer Engineering, Wayne State University, Detroit, MI, USA
[5]Department of Physics and Astronomy, Wayne State University, Detroit, MI, USA
[6]College of Integrative Sciences and Arts, Arizona State University, Mesa, AZ, USA
[7]Department of Physics, Arizona State University, Tempe, AZ, USA

We combine modeling and experiments to investigate second- and third-harmonic generation (SHG/THG) in metal-indium tin oxide (ITO) metasurfaces. Linear optics at normal incidence show moderate field enhancement near the ITO epsilon-near-zero (ENZ) wavelength, steering the focus toward intrinsic, material driven nonlinear response rather than simple linear field boosting. Wavelength resolved SHG requires a Lorentz-dispersive $\chi^{(2)}$ for ITO to match spectra; a static $\chi^{(2)}$ fails. Angle-resolved SHG/THG cannot be reproduced with purely real coefficients; grouped contributions to $\chi^{(3)}$ (and effective $\chi^{(2)}$) must be complex. Using a hydrodynamic model for the metal and ITO with linear dispersion plus dispersive $\chi^{(2)}$ and $\chi^{(3)}$, we show that these complex phases arise from coherent interference of nonlinear sources in the metal, ITO, and interfaces, each weighted by distinct, complex local field and radiation factors. Experimentally, we fabricated split-ring resonator metasurfaces on ITO films atop a metallic ground plate and measured linear reflectance and angle-resolved SHG/THG in reflection geometry. The measurements quantitatively confirm the modeling: dispersive $\chi^{(2)}$ is necessary to capture SHG spectra, and complex, interference-induced effective coefficients are essential to reproduce angular SHG/THG patterns. Together, these results provide a unified, physically grounded interpretation of nonlinear emission from metal-oxide metasurfaces without relying on ENZ field enhancement.

**Introduction**

Nonlinear metasurfaces enable compact, phase-programmable frequency conversion by engineering resonant near fields and radiation channels at the subwavelength scale.[1] By concentrating optical energy in nanoscale volumes and tailoring symmetry, these structures have delivered efficient second-harmonic (SHG)[2–4] and third-harmonic generation (THG),[5] polarization control,[6] and beam shaping, with applications spanning coherent sources, spectroscopy, and on-chip photonics.[7] A particularly promising materials platform leverages epsilon-near-zero (ENZ) media[8–10] (such as indium tin oxide (ITO)[11]) with real pat of permittivity crossing zero in the near-infrared. Near the ENZ point, electric field component normal to the film is enhanced, nonlinear source terms are boosted by local intensity, and small spectral changes yield large, tunable nonlinear responses.[12]

Metal-dielectric-metal stacks that embed ENZ layers beneath plasmonic resonators[13] combine multiple advantages: strong capacitive confinement in split-ring resonators (SRRs), impedance-matched outcoupling via the ground plane, and thickness-controlled phase accumulation in the spacer. This architecture supports multipath nonlinear radiation: "same-axis" pathways where the same field component drives and radiates, mixed pathways that couple orthogonal components through anisotropy or symmetry breaking, and interfacial versus bulk sources that accumulate distinct propagation phases before emission. Understanding which pathways dominate, and how their complex phases interfere across angle, is crucial for predicting and designing the angular fingerprints of SHG and THG.

Accurate modeling of such systems requires going beyond local, dispersionless approximations. For noble metals, the hydrodynamic electron model captures nonlocal screening, convective nonlinearities, and the interplay between interband damping and free-electron response under strong confinement.[14–20] For conducting oxides like ITO, both the linear permittivity and the effective $\chi^{(2)}$, $\chi^{(3)}$ are markedly dispersive near ENZ,[21] reflecting resonant bound-charge and free-carrier contributions. Treating $\chi^{(2)}$ and $\chi^{(3)}$ as static parameters can misrepresent the magnitude and the phase of the nonlinear polarization, thereby mispredicting interference between pathways and the angular properties of harmonics in the far-field.[22]

Recent works by Scalora and co-workers[23,24] has established a rigorous hydrodynamic framework for the electrodynamics of conductive oxides, with ITO as a central case study. Ref. [23] combined spectral and angular SHG/THG measurements from ultrathin ITO with a phenomenological hydrodynamic model that treats free- and bound-electron polarizations, nonlocal pressure/viscosity, magnetic (Lorentz) terms, surface/bulk sources, and hot-electron–induced mass changes. It was shown that a comprehensive material model is necessary to reproduce measured SHG/THG without invoking ad hoc, dispersionless nonlinearities. Related experimental reports from the same community on ITO thin films near ENZ similarly emphasized that accurate predictions require microscopic, time-domain models that preserve linear and nonlinear dispersion and the phase of the sources, rather than attributing enhancement to the ENZ crossing alone.

Building on these insights, we target a different physical regime and architecture: periodic split-ring resonator (SRR)/ITO/Au metasurfaces operating in reflection. Here, multipolar SRR resonances,[25,26] capacitive gaps,[27] and the metal back-plane create strong, symmetry-broken near fields and well-defined radiation channels. In this setting, both the magnitude and the phase of ITO's internal nonlinearities governs interference between same-axis and mixed pathways, which in turn determines the angular fingerprints of SHG and THG.

To capture this physics, we couple a self-consistent hydrodynamic description of the metals to an in-house FDTD solver and represent ITO with three increasing levels of fidelity: (1) linear dispersive dielectric; (2) addition of static $\chi^{(2)}$ and $\chi^{(3)}$; and (3) addition of Lorentz-dispersive $\chi^{(2)}(\omega;2\omega)$ and $\chi^{(3)}(\omega;3\omega)$. In parallel, we derive a compact driven oscillator model that yields closed-form angular dependences for SHG and THG; by fitting angular scans with complex coefficients, it encodes the amplitude and phase of distinct radiation pathways in an interpretable manner.

This combined numerical-analytical framework reveals that dispersive $\chi^{(2)}$ and $\chi^{(3)}$ of ITO are required to reproduce the measured harmonic intensities and their angular structure. The Lorentz-dispersive model achieves excellent agreement with experiment for SHG, whereas frequency-independent nonlinearities fail to capture lobe orientations, node depths, and contrast. The analytical fits trace these features to phase differences among sources, especially mixed terms that are driven into quadrature with same-axis contributions by ENZ-induced dispersion and propagation through the multilayer stack. Thus, in resonant SRR/ITO/Au metasurfaces, enhanced harmonic generation arises from the interplay between SRR induced modal fields and intrinsically dispersive ITO nonlinearities, rather than from ENZ field concentration alone. This perspective reconciles experiment and simulation and furnishes a practical route to inverse design: by targeting desired angular signatures and yields, one can solve for the complex pathway coefficients and map them to structural and material parameters (such as ITO thickness, SRR geometry, and dispersion engineering) to realize angle-tailored nonlinear emission.

**Experimental Platform and Measurement Methods**

The metasurface comprises gold split-ring resonators (SRRs) patterned on an indium-tin-oxide (ITO) film deposited over a continuous gold ground plate on a dielectric substrate (Fig. 1b). This metal-ITO-metal stack suppresses transmission and enforces reflection-only operation for both the pump and the generated harmonics. ITO films, with carrier concentration and mobility chosen to place the epsilon-near-zero (ENZ) region near the target band around 1615 nm, were deposited by a DC magnetron reactive sputtering system on an Au ground plate. The ground plate was formed by e-beam–evaporated Au of thickness $L_{Au}$ = 200 nm with a Ti adhesion layer of 5 nm, and the ITO thickness was $L_{ITO}$ = 85 nm. SRRs were defined by electron-beam lithography using ZEP 520A resist, followed by Au deposition to a thickness $L_z$ = 75 nm and lift-off. The array period 396×396 nm$^2$ was chosen to avoid diffractive orders in the wavelength and angle ranges used. Nominal in-plane dimensions are $L_x$ = 264 nm, $L_y$ = 236 nm, and the width w = 71.4 nm, consistent with the representative SEM in Fig. 1c. Post-fabrication, SEM was used for dimensional verification and uniformity.

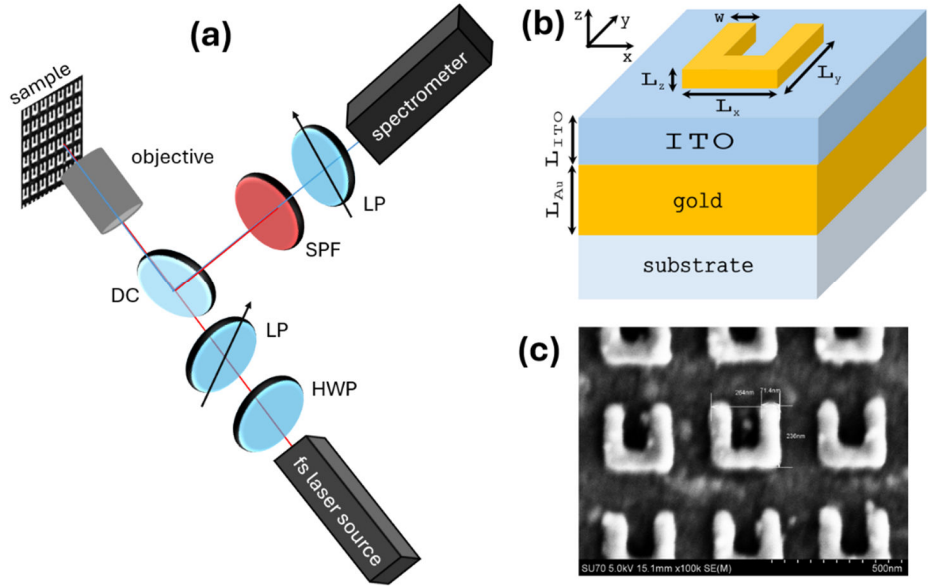

*Figure 1*. Experimental setup and sample architecture. (a) The reflection-geometry for nonlinear microscope used for SHG/THG measurements. Linear measurements were performed with a commercial FTIR. A femtosecond laser is polarization-conditioned with a half-wave plate (HWP) and linear polarizer (LP), directed through a dichroic mirror (DC), and focused onto the metasurface with a 10x, 0.26 NA objective. The generated harmonics are collected by the same objective, pass a short-pass filter (SPF) to block the fundamental, cleaned by an analyzing LP, and are sent to a spectrometer. (b) Schematics of the metasurface: gold split-ring resonators (SRRs) patterned on an indium–tin-oxide (ITO) film deposited atop a gold ground plate on a substrate. Geometric parameters are indicated: SRR outer lengths $L_x$=264 nm, $L_y$=236 nm, width w=71.4 nm, thickness $L_z$=75 nm, and layer thicknesses $L_{ITO}$=85 nm and $L_{Au}$=200 nm. (c) the SEM image of the fabricated SRR array showing uniform U-shaped resonators; representative in-plane dimensions are annotated (scale bar: 500 nm).

Normal-incidence reflectance spectra were measured with a broadband source and a microscope coupled Fourier-transform infrared spectrometer (Bruker Vertex 80V) using polarization control and referencing to a gold mirror for absolute reflectance. These measurements confirm minimal field enhancement near the ITO ENZ wavelength at normal incidence and supply inputs to the electromagnetic model, including layer thicknesses, effective indices, and loss.

Second- and third-harmonic signals were recorded in a reflection-geometry microscope (Fig. 1a). A femtosecond laser, tunable from 1.0 – 2.0 micron with pulse duration of 120 fs and repetition rate of 80 MHz, was passed through a half-wave plate and linear polarizer for input polarization control and power stabilization, routed via a dichroic mirror, and focused onto the sample by a 0.26 NA objective to a spot size of ~1 μm. The generated harmonics were epi-collected by the same objective, filtered by a short-pass to reject the fundamental, analyzed by an output polarizer to select *s* or *p* components, and delivered to a calibrated spectrometer. For narrowband acquisition we inserted interference filters centered at 2ω and 3ω. Pump powers and integration times were chosen to remain well below the damage threshold and within the perturbative regime; power-scaling tests verified quadratic (SHG) and cubic (THG) dependencies.

Angle-resolved datasets were acquired at normal incidence by rotating the in-plane pump polarization with a motorized polarization plate; the sample and collection optics remained fixed.

For each spectrum we recorded the analyzer setting (x/y or arbitrary linear basis) along with the pump polarization angle. No goniometric scans were used for linear or nonlinear measurements. The effective collection solid angle and the wavelength-dependent transmission of all optical elements were calibrated and incorporated into the radiation/collection model to enable quantitative comparison with simulations.

Background signals were evaluated on unpatterned regions and subtracted from the data; the Au ground plate without SRRs produced negligible SHG/THG under identical conditions. We checked for multiphoton photoluminescence artifacts by inspecting spectra for broadband tails and by confirming perturbative power scaling. Polarization-selective measurements (various in-out combinations) were used to isolate dominant tensor channels and to benchmark modeled radiation patterns. All reported intensities are corrected for detector and grating response, and uncertainties reflect repeat measurements across multiple array locations. The experimentally determined reflectance, measured SRR dimensions, and analyzer settings serve as direct inputs to the modeling framework used in the subsequent sections.

**Modeling Framework**

Optical properties of the metasurface depicted in Fig. 1 are simulated with an in-house finite-difference time-domain (FDTD) solver.[28] Electromagnetic fields are propagated in real space and time by integrating Maxwell's equations,

$$\dot{\mathbf{B}} = -\nabla \times \mathbf{E},$$
$$\dot{\mathbf{E}} = c^2 \nabla \times \mathbf{B} - \frac{1}{\varepsilon_0}\dot{\mathbf{P}}, \qquad (1)$$

where $\mathbf{E}$ and $\mathbf{B}$ are the electric and magnetic fields, respectively, and $\dot{\mathbf{P}} \equiv \mathbf{J}$ the total material current density. To describe the metal, we adopt a Drude-Lorentz decomposition and then extend the Drude channel to the nonlinear hydrodynamic regime. The total current in the metal is written as

$$\mathbf{J} = \mathbf{J}_D + \mathbf{J}_L, \qquad (2)$$

where $\mathbf{J}_D$ and $\mathbf{J}_L$ are the current densities following the Drude (free-electron) and the Lorentz (bound-electron) contributions, respectively, with $\dot{\mathbf{P}}_D \equiv \mathbf{J}_D$ and $\dot{\mathbf{P}}_L \equiv \mathbf{J}_L$. The linear response of gold is parameterized using the oscillator set as in Ref. [29].

Nonlinear effects in the free-electron gas are modeled using the semiclassical hydrodynamic approach[20,30–32]. The Drude polarization, $\mathbf{P}_D$, obeys

$$\ddot{\mathbf{P}}_D + \gamma\dot{\mathbf{P}}_D = \varepsilon_0\omega_p^2\mathbf{E} + \frac{e}{m_e^*}\left(\dot{\mathbf{P}}_D \times \mathbf{B} - (\nabla\cdot\mathbf{P}_D)\mathbf{E}\right) - \frac{1}{n_0 e}\left((\nabla\cdot\dot{\mathbf{P}}_D)\dot{\mathbf{P}}_D + (\nabla\cdot\dot{\mathbf{P}}_D)\dot{\mathbf{P}}_D + \varepsilon_0\omega_p^2\nabla p\right), \qquad (3)$$

where $\gamma$ is the momentum-relaxation rate, $\omega_p$ is the plasma frequency, $m_e^*$ is the effective electron mass. The final term represents the pressure of the electron gas and is written following Ref. [33] as

$$p = n_0 E_F \nabla \left(\frac{n}{n_0}\right)^{5/3} \quad (4)$$

with $n$ being the carrier density, $n_0$ is its equilibrium value, and $E_F$ is the Fermi energy. In practice we expand the pressure contribution to third order in $\mathbf{P}_D$,[32] which yields the nonlocal "pressure" correction at leading order together with the second- and third-order source terms that drive SHG and THG. The Lorentz (bound-electron) oscillators are kept strictly linear and serve only to reproduce the interband dispersion of gold.

This Drude-only nonlinear pole provides a minimal yet physically grounded description that captures the observed harmonic generation without invoking phenomenological nonlinearities of the bound electrons. Although such terms can be added if needed (see, e.g., Ref. [16]), we find that the present model explains the measurements and enables consistent comparison between experiment and simulation. In all simulations, we enforce the normal component of $\mathbf{J}_D$ at the metal boundaries (hard-wall condition) and couple the auxiliary equations for $\mathbf{P}_D$ to the FDTD leap-frog update through standard auxiliary differential equations (ADE) time stepping.

We use plasma frequency $\omega_P$ and damping rate $\gamma$ from Ref. [29] for the linear Drude pole. To ensure consistency with the hydrodynamic model, we fix the conduction-electron density to the tabulated value for gold ($n_0 = 5.90 \times 10^{28}$ m$^{-3}$) and solve $\omega_p^2 = e^2 n_0 / (\varepsilon_0 m_e^*)$ for the effective mass, obtaining $m_e^* = 1.31 m_e$. The value of the Fermi energy, $E_F$, is 5.53 eV. This procedure preserves the linear response while providing self-consistent parameters for the nonlinear hydrodynamic terms.

We model the ITO spacer with three progressively richer descriptions that separate linear dispersion from nonlinear sources while remaining compatible with time-domain simulation. First, the linear permittivity $\varepsilon(\omega)$ is described by a Drude-Lorentz fit to spectroscopic ellipsometry data taken from[34] with a slight adjustment of the parameter $\varepsilon_s$ to shift the ENZ wavelength to 1615 nm. This representation captures the free-carrier response that is responsible for the ENZ region as well as interband contributions at higher energies. In the FDTD solver, the Drude pole and a single Lorentz oscillator are implemented via ADE approach (same as the one used for the Drude-Lorentz response of gold) so that the time-domain updates reproduce the target $\varepsilon(\omega)$ across the spectral window of interest. The same $\varepsilon(\omega)$ is used in all nonlinear scenarios to ensure that any changes in harmonic generation arise from the nonlinear sources alone.

**Table I**

|  | $\chi^{(n)}$, static | $B_n$ | $\Omega_n$ | $\gamma_n$ |
|---|---|---|---|---|
| n = 2 | $2 \times 10^{-12}$ m/V | $6.67 \times 10^{17}$ m rad$^2$/(V sec$^2$) | $1.167 \times 10^{15}$ rad/sec | $1.45 \times 10^{14}$ rad/sec |
| n = 3 | $10^{-21}$ (m/V)$^2$ | $5.07 \times 10^9$ (m rad/(V sec))$^2$ | $1.167 \times 10^{15}$ rad/sec | $1.45 \times 10^{14}$ rad/sec |

Parameters for nonlinear simulations of ITO.

Second, on top of the linear response function, we consider two types of nonlinear description of ITO: constant, frequency independent susceptibilities and dispersive susceptibilities governed by the Lorentzian response. As a baseline, we adopt an isotropic third-order susceptibility $\chi^{(3)}$ and an effective second-order susceptibility $\chi^{(2)}$ within the ITO layer. The $\chi^{(3)}$ polarization is taken in the standard instantaneous scalar-Kerr form

$$\tilde{\mathbf{P}}_{\mathrm{nl}}^{(3)} = \varepsilon_0 \chi^{(3)} \mathbf{E}(\mathbf{E}\cdot\mathbf{E}), \tag{5}$$

which respects permutation symmetry and captures self- and cross-phase modulation and the source for THG. For $\chi^{(2)}$, we use an effective form aligned with the layer normal to z that reflects inversion-symmetry breaking

$$\begin{aligned}
\tilde{P}_{\mathrm{nl,x}}^{(2)} &= 4\varepsilon_0 \chi^{(2)} E_x E_z, \\
\tilde{P}_{\mathrm{nl,y}}^{(2)} &= 4\varepsilon_0 \chi^{(2)} E_y E_z, \\
\tilde{P}_{\mathrm{nl,z}}^{(2)} &= 2\varepsilon_0 \chi^{(2)} (\mathbf{E}\cdot\mathbf{E}).
\end{aligned} \tag{6}$$

This tensorial structure enforces the expected selection rules for even order processes in a film bounded by different optical media and allows us to encode the dominant out-of-plane symmetry breaking without invoking a full microscopic interface model.

To allow for spectral structure in the nonlinearities (particularly relevant near the ENZ region where both the local density of states and the field distribution vary rapidly) we also implement dispersive $\chi^{(2)}(\omega)$ and $\chi^{(3)}(\omega)$. Each susceptibility is represented as a single-pole Lorentzian with distinct amplitudes and resonance parameters

$$\chi^{(n)}(\omega) = \frac{B_n}{\Omega_n^2 - \omega^2 - i\gamma_n \omega}. \tag{7}$$

In the time domain, this choice translates into driven Lorentz equations for the corresponding nonlinear polarizations

$$\begin{aligned}
&\ddot{\mathbf{P}}_{\mathrm{nl}}^{(n)} + \gamma_n \dot{\mathbf{P}}_{\mathrm{nl}}^{(n)} + \Omega_n^2 \mathbf{P}_{\mathrm{nl}}^{(n)} = \varepsilon_0 B_n \mathbf{F}^{(n)}, \\
&\left(\mathbf{F}^{(2)}\right)_x = 4 E_x E_z, \left(\mathbf{F}^{(2)}\right)_y = 4 E_y E_z, \left(\mathbf{F}^{(2)}\right)_z = 2(\mathbf{E}\cdot\mathbf{E}), \\
&\mathbf{F}^{(3)} = \mathbf{E}(\mathbf{E}\cdot\mathbf{E}).
\end{aligned} \tag{8}$$

While this construction is causal, it provides a minimal parametric handle to describe enhancement or suppression of the nonlinear response around specific frequencies, and it remains numerically stable in FDTD because the auxiliary variables are updated alongside the Maxwell fields. The numerical parameters used in all nonlinear models for ITO are listed in Table I.

A key modeling decision concerns the spatial distribution and the physical origin of $\chi^{(2)}$ in ITO. In an ideal centrosymmetric bulk, the $\chi^{(2)}$ terms vanish. Nevertheless, SHG from ITO films is widely observed because several mechanisms may break inversion symmetry.[35] The most familiar is interfacial symmetry breaking: at the metal/ITO and ITO/dielectric boundaries, the termination of the lattice and the abrupt change of the electronic environment produce an even order response that can be modeled as an ultrathin sheet of nonlinear polarization.[36] A second mechanism is the nonlocal bulk contribution,[37] which originates from spatial dispersion of the electron polarization and magnetic-dipole terms. These sources are allowed in centrosymmetric media and scale with field gradients,[38] which can be significant in nanostructures and near the ENZ where the longitudinal field and its divergence are enhanced.

A third and often dominant mechanism in conducting oxides is electric-field-induced second-harmonic generation (EFISH),[39] in which a third-order nonlinearity mixes with a static or slowly varying internal field, $E_{DC}$, to produce an effective $\chi^{(2)} \approx 3 \chi^{(3)} E_{DC}$. In degenerately doped oxides, built-in electrostatic fields arise naturally from band bending at metal/oxide and oxide/dielectric interfaces, from depletion or accumulation layers set by work-function mismatch, from trapped charges, and from stoichiometry gradients.[40] The magnitude and sign of $E_{DC}$ can vary across only a few nanometers, but because $\chi^{(3)}$ in ITO is large and the ENZ condition amplifies the local field, the resulting $\chi^{(2)}_{\text{eff}}$ can be substantial. EFISH corresponds to the $\chi^{(3)}(\omega; \omega, 0, \omega)$ mixing channel, where a quasi-static or low-frequency field couples with the optical field to yield an effective $\chi^{(2)}$. Although our experimental setup does not explicitly have any bias applied, one possibility to generating a low-frequency field under ultrafast excitation is difference-frequency generation (DFG) in the vicinity of strong field enhancement, for example near localized plasmon resonances of SRRs.[41] In that case, the ($\omega - \omega`$) components of the nonlinear polarization can produce a low-frequency or near-DC polarization[42] that follows the pulse envelope, which during the pulse acts as an effective $E_{DC}$ and can drive EFISH-type SHG. The key requirement is that the DFG component be sufficiently strong and slowly varying compared to the optical period so that its mixing with the fundamental is phase-coherent over the interaction length. Under intense femtosecond excitation in plasmonic environments, such rectified fields can reach appreciable amplitudes due to large local-field factors, potentially making EFISH relevant even without an externally applied bias. We note that THz induced EFISH has been recently observed experimentally in ITO.[43]

In the context of experiments performed in the optical region of 1 to 2 microns, this mechanism can plausibly be present if our geometry supports efficient DFG in regions of strong field confinement near SRR/ITO interfaces. At the ENZ wavelength, longitudinal fields inside ITO are enhanced and spatial gradients are large, which can further boost both the efficiency of low-frequency generation and its ability to act uniformly across the thin layer. If present, the resulting low frequency field would be synchronized with the pump envelope, effectively quasi-static over many optical cycles during the pulse, and, therefore, capable of contributing to SHG via EFISH. We emphasize, however, that in our measurements we do not observe independent signatures that require invoking such a mechanism; the observed SHG can be captured by a homogenized $\chi^{(2)}$ representing interfacial symmetry breaking and nonlocal bulk sources. We, therefore, regard DFG-assisted EFISH as a plausible but unverified contributor that could be tested in future work by probing repetition-rate and average-power dependencies at fixed peak intensity, by time resolving the SHG build-up within and after the pulse, or by engineering plasmonic features to deliberately enhance the low-frequency rectification and assessing the correlated change in SHG.

Lastly, our ITO layer is deeply subwavelength at both the pump and harmonic frequencies. In this thin film limit, the detailed axial distribution of even order sources (surface sheets at both interfaces, EFISH concentrated in space-charge regions, and nonlocal bulk terms tied to field gradients) produces second harmonic fields that add with phases determined largely by the pump field profile rather than by the exact nanometer scale position of each source. Consequently, a homogenized description in which an effective $\chi^{(2)}$ is uniformly distributed throughout the ITO layer provides an accurate surrogate for the aggregate even-order response. This volumetric $\chi^{(2)}$ reproduces the observed SHG efficiencies and polarization dependences within experimental uncertainty while avoiding underconstrained assumptions about which interface or mechanism dominates. From a computational perspective, the volumetric representation also circumvents the need for sub-cell surface sources or discontinuity conditions, which can be delicate in FDTD when strong ENZ-enhanced fields generate steep spatial gradients.

The dispersive $\chi^{(n)}(\omega)$ model (7) allows us to test whether the nonlinear coefficients are approximately constant across our tuning range or exhibit resonant enhancement. Although our data are adequately described with one Lorentz pole, additional structure could be incorporated by adding poles if warranted by future measurements. There are, of course, limits to the homogenized $\chi^{(2)}$ approximation. It does not by itself distinguish the relative phases and magnitudes of the metal/ITO and ITO/dielectric interface contributions. If future experiments reveal sensitivity to sample orientation, applied bias, or layer thickness that cannot be reconciled within the uniform model, the framework can be extended to include coordinate dependent $\chi^{(2)}$ tensors.

**Results and Discussion**

We first address the system's linear response. Figure 2 presents the experimental and theoretical spectra and the corresponding intensity profiles from our linear model. The good agreement between the experiment and our model shown in panels (a) and (b) establishes that the linear response of the stack is captured reliably: the x-polarized SRR resonance is purposefully red-shifted by the ground plate and by the ITO spacer to lie near the ITO ENZ crossing, while the orthogonal polarization excites a weaker, higher frequency mode. Yet the field maps and line profiles in panels (c) – (e) show that the local intensity inside ITO remains modest. This outcome reflects how longitudinal fields and phase relations are constrained in our geometry. At normal incidence the incident wave carries no in-plane wavevector, so the only path to a strong longitudinal $E_z$ in the spacer is through near-field conversion by the resonator. The SRR does supply some $E_z$ at metal edges and across the gap, but in the presence of a continuous Au back reflector the ITO functions as the dielectric core of a metal-insulator-metal cavity whose thickness is far below a quarter wavelength. The standing wave condition fixed by the mirror enforces a particular phase of the reflected field; for our thickness this places the longitudinal field maximum slightly above the spacer rather than deep inside it, as evident from panel (e). Consequently, the ENZ condition, while it reduces the real part of $\varepsilon$ and can, in principle, favor longitudinal displacements,[44] does not translate here into a high-Q Berreman-like absorption with giant in-spacer $E_z$.[45]

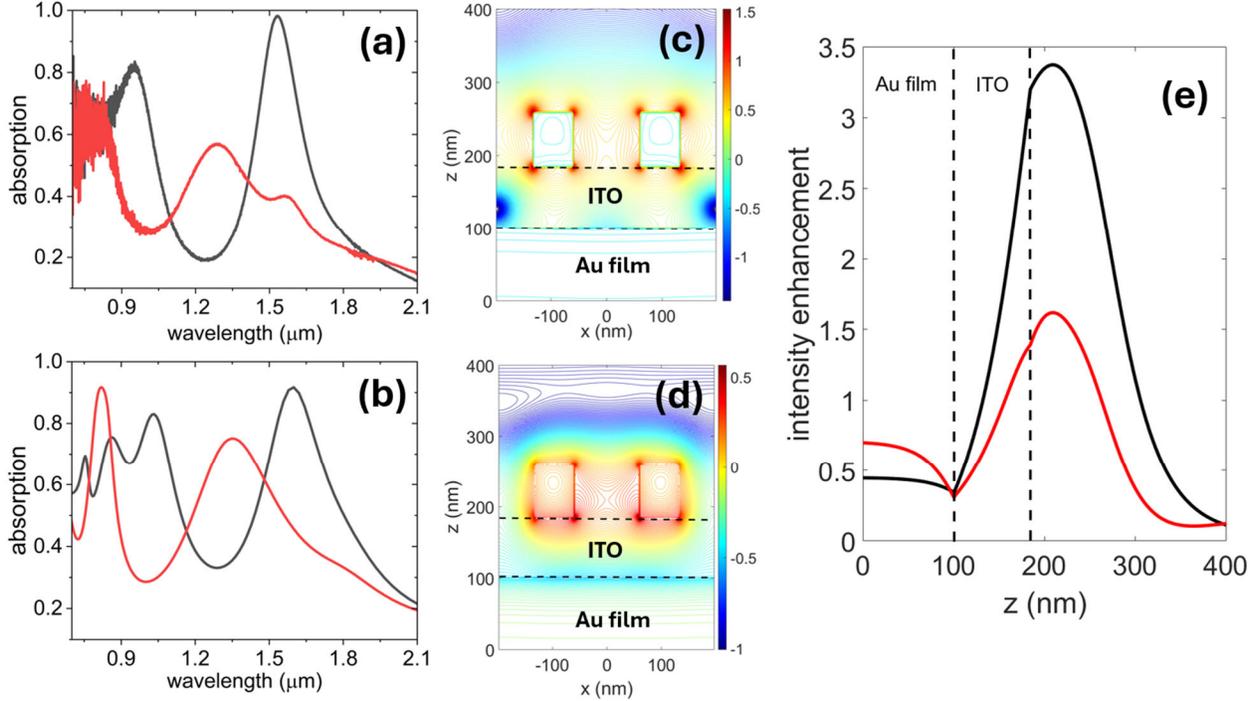

*Figure 2*. Linear optical response of the SRR/ITO/Au stack at normal incidence. (a) Measured absorption A = 1 − R for x-polarized (black) and y-polarized (red) probes. (b) Corresponding FDTD results using the linear Drude–Lorentz ITO model. The X-polarized resonance is aligned with the ITO ENZ wavelength near 1.6 μm; the Y-polarized response peaks near 1.3 μm. (c,d) FDTD xz cuts (y = 0) of the intensity enhancement on a logarithmic scale at 1615 nm (x-pol, c) and 1300 nm (y-pol, d). Dashed lines mark the ITO spacer and Au ground film. (e) Linear-scale intensity enhancement versus z at x = 0 extracted from (c,d); black: X-pol at 1615 nm, red: Y-pol at 1300 nm. The excitation impinges from positive z.

Importantly, in this configuration strong SHG and THG are observed near the ENZ wavelength (see Fig. 3 below) even though the local ENZ-driven field enhancement in the linear regime is only moderate. The full multilayer/metasurface structure, together with the SRR resonance, provides efficient channels for frequency conversion – through interfacial symmetry breaking, nonlocal longitudinal response in ITO near ENZ, and dispersive $\chi^{(2)}$ and $\chi^{(3)}$ contributions (without relying on extreme hot-spot amplification). Normal incidence excitation together with the ground plate red-shifts the SRR into the ENZ region but limits the build-up of a large uniform $E_z$ within the spacer,[46] so the intensity inside ITO remains modest and well characterized. This separation between conversion efficiency and local-field magnitude allows the measured SHG and THG to be attributed predominantly to the intrinsic, ENZ-sensitive nonlinearities of ITO and its interfaces rather than to simple cavity-driven enhancement.

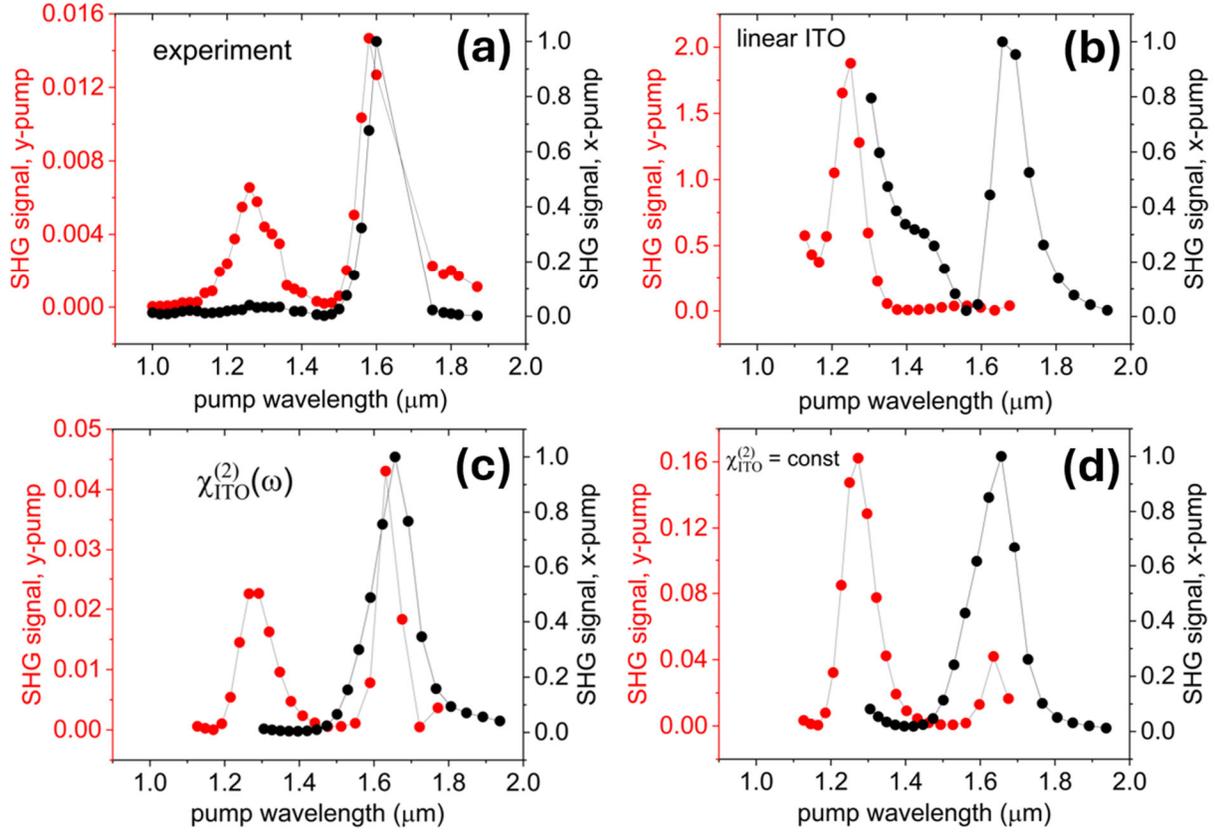

*Figure 3*. Total SHG versus pump wavelength for x-polarized (black circles, right axis) and y-polarized (red circles, left axis) excitation. In all panels, both traces are normalized to the maximum SHG signal obtained for x-polarized pumping. (a) Experiment. All simulations on (b – d) incorporate hydrodynamics model for metal. (b) Simulations with linear, dispersive ITO (Drude–Lorentz; no $\chi^{(2)}$ in ITO). (c) Simulations with a dispersive $\chi^{(2)}$ for ITO. (d) Simulations with a constant $\chi^{(2)}$ for ITO. Only the model with dispersive $\chi^{(2)}$ in ITO reproduces the experimental spectral shape and polarization contrast.

Figure 3 compares polarization resolved SHG spectra from experiment with three modeling levels to identify which ingredients are necessary to reproduce the data. Panel (a) benchmarks the measurement. Two clear SHG bands are observed: a y-pump feature around 1.25 – 1.3 µm and a pronounced x-pump resonance at 1.6 – 1.65 µm with comparatively low output elsewhere. The dual vertical axes allow a direct comparison between polarization channels after normalizing the data to the maximum SHG obtained from the x-pump for each set of results; this highlights relative spectral shapes and polarization contrasts rather than absolute conversion efficiencies.

All simulations employ the hydrodynamic description of the metal (3). Panel (b) considers a realistic linear dispersion for ITO, but it excludes any intrinsic second order response from ITO. In this scenario, SHG originates only from metallic multipolar sources and interface dipoles. While some spectral features appear near the resonant wavelengths of the structure, the overall behavior significantly deviates from the experimental observations: the short wavelength y-pump response is overestimated and the polarization balance around 1.6 µm is incorrect. This indicates that metal dominated sources alone cannot reproduce the measured polarization resolved spectra.

Despite the discrepancy, it is interesting to note that in panel (b) the x-pump trace (black circles) exhibits a clear Fano profile: a dispersive, asymmetric lineshape with a pronounced peak-dip sequence across the resonance. This arises from coherent interference among multiple SH pathways with different phase dispersion. One is the "broad" background generated at the metal ground plate and other nonresonant interfaces. A second is the "narrow" channel associated with the SRR mediated SH source, whose amplitude and phase vary rapidly with pump wavelength due to the underlying fundamental resonance and the hydrodynamic nonlocal response in the metal. A third, closely related pathway is the SRR generated SH that reflects off the ground plate and re-radiates; the ground plate thus acts as a mirror that feeds back a delayed replica of the resonant SRR SH field. As the pump scans through the SRR resonance, the phase of the resonant SH channel(s) winds by roughly $\pi$ relative to the broadband background, while the reflected SRR SH acquires an additional phase set by twice the spacer optical thickness. The superposition of these components produces constructive interference on one side of resonance (the peak) and destructive interference on the other (the dip/zero crossing). The reflective ground plate is therefore doubly important: it supplies a strong, spectrally smooth SH reference and also returns a phase-shifted copy of the SRR generated SH. Consequently, the asymmetry and depth of the dip encode the amplitude ratios and relative phases among (i) the ground plate/background SH, (ii) the direct SRR SH, and (iii) the ground plate reflected SRR SH.

Panel (c) adds a frequency dependent $\chi^{(2)}$ for the ITO spacer. The simulations closely track experiment for both polarizations: it captures the position, relative height, and width of the y-pump band near 1.25-1.3 µm and the strong, narrow x-pump peak around 1.6 µm. The agreement implies that the spacer behaves as an active second order medium with effective $\chi^{(2)}$ changing significantly with wavelength. To achieve this level of agreement, we varied the parameters of a Lorentzian model for $\chi^{(2)}(\omega)$ (7) and identified an optimal set through extensive simulations; the resulting best fit parameters are reported in Table I. These optimized values govern both the magnitude and phase dispersion required to reproduce the observed peak positions, widths, and relative amplitudes across polarizations. The dispersion of this response is consistent with the broader physical picture developed earlier, in which multiple microscopic channels (surface symmetry breaking at interfaces, nonlocal bulk (gradient and magnetic-dipole) terms, carrier-related contributions, and structure-dependent longitudinal fields) combine with resonant field enhancement to yield a strongly dispersive magnitude and phase of $\chi^{(2)}$.

Panel (d) keeps an ITO $\chi^{(2)}$ but forces it to be frequency independent. Although this reproduces the general presence of two bands set by the linear resonances of the stack, it fails to match the measured polarization contrast and relative amplitudes: the short wavelength y-pump band is too strong and broad, and the 1.6 µm region lacks the correct balance between x- and y-pump responses. This shows that the wavelength dependence of both the magnitude and the phase of the spacer's effective $\chi^{(2)}$ is essential.

Taken together, the four panels lead to three conclusions. First, the resonant modal structure of the metasurface determines where fields and field gradients are large enough to drive SHG. Second, metallic surface and multipolar sources alone are insufficient to account for the spectra.

Third, the ITO spacer provides a dispersive second order response. It arises from the combined, wavelength dependent influence of the mechanisms discussed previously, which is required to reproduce the observed spectral selectivity and polarization dependence.

This strategy leverages the tighter spectral constraints available from SHG to set THG parameters in order to match the angular data (Fig. 4). While the lack of THG spectra introduces some non-uniqueness in $\chi^{(3)}$, the resulting parameter set (Table I) is physically plausible, consistent with the SHG-validated dispersion, and sufficient to capture the salient features of the THG angular response. We view full THG spectral measurements as a valuable future step to further refine the $\chi^{(3)}$ dispersion; however, the present assumptions do not affect our central conclusion that dispersive nonlinearities in ITO and their phases are required to explain the observed SHG/THG angular fingerprints in SRR/ITO/metal metasurfaces.

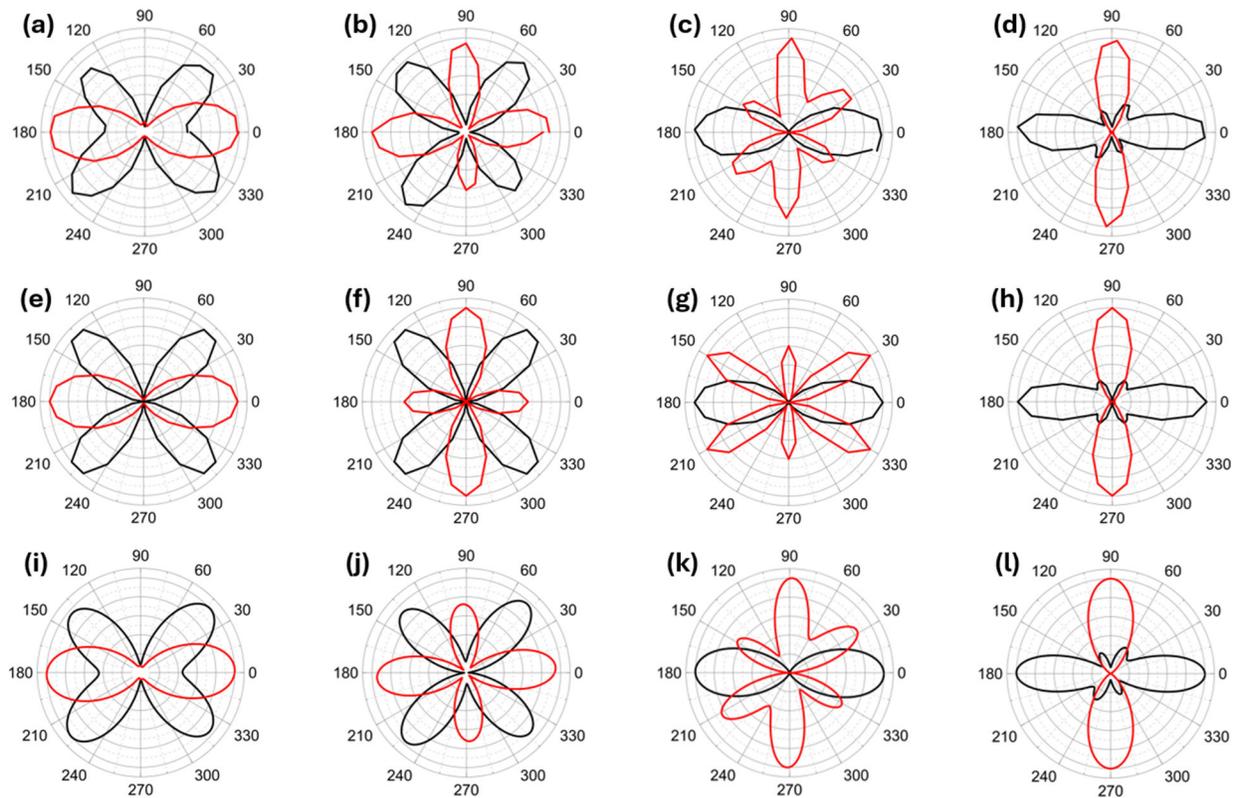

*Figure 4*. Angular resolved SHG and THG. Polar plots (normalized) show x-polarized (black) and y-polarized (red) harmonic outputs as functions of the in-plane pump polarization angle $\theta$ (degrees). Top row, experiment: (a) SHG at $\lambda_{pump}$ = 1615 nm; (b) SHG at 1300 nm; (c) THG at 1615 nm; (d) THG at 1300 nm. Middle row, the hydrodynamic model for metal and dispersive $\chi^{(2)}$ and $\chi^{(3)}$ for ITO: (e-h) same pump wavelengths and panel order as in (a-d). Bottom row, analytical-model fits: (i-l) fitted angular responses corresponding to (a-d); fitting parameters are listed in Table II.

Angularly resolved SHG and THG polar patterns provide a direct diagnostic of which in-plane components of the nonlinear susceptibility contribute to an overall nonlinear signal at a given pump wavelength. In the configuration of Fig. 1b, the SRRs are oriented with x along the gap axis and y being transverse. For the purposes of interpreting Fig. 4, we restrict attention to in-plane driving fields and in-plane detected signals, so only $\chi^{(2)}$ and $\chi^{(3)}$ components with x and y indices are relevant.

The second order response reads as follows. The far-field SHG detected in x or y polarization arises from in-plane contractions of the form

$$P_x(2\omega) \sim \chi^{(2)}_{xxx} \cos^2(\theta) + \chi^{(2)}_{xxy} \sin(2\theta) + \chi^{(2)}_{xyy} \sin^2(\theta),$$
$$P_y(2\omega) \sim \chi^{(2)}_{yxx} \cos^2(\theta) + \chi^{(2)}_{yxy} \sin(2\theta) + \chi^{(2)}_{yyy} \sin^2(\theta), \tag{9}$$

where $\theta$ is the polar angle in xy-plane of the SRR. Here the effective contributions are governed by the in-plane sets $\{\chi^{(2)}_{xxx}, \chi^{(2)}_{xxy} = \chi^{(2)}_{xyx}, \chi^{(2)}_{xyy}\}$ for x-polarized SHG and $\{\chi^{(2)}_{yxx}, \chi^{(2)}_{yxy} = \chi^{(2)}_{yyx}, \chi^{(2)}_{yyy}\}$ for y-polarized SHG. The SRR symmetry imposes anisotropic coupling between the gap-aligned and transverse directions, so these coefficients need not be equal when x and y are interchanged. As the pump polarization is rotated in the xy plane, different quadratic field products $E_i \cdot E_j$ dominate, which produces the two-lobe SHG patterns and their angular shifts between x and y detection channels seen in Fig. 4. In practice, $\chi^{(2)}$ elements that mix x and y (e.g., xxy/xyx and yxy/yyx) govern the rotation of the lobes and the relative contrast between the x- and y-polarized SHG intensities.

For the third order, the relevant independent in-plane components can be grouped by index permutations: for x-polarized THG, $\chi^{(3)}_{xxxx}$ captures the response to $E_x^3$, while $\{\chi^{(3)}_{xxxy}, \chi^{(3)}_{xxyx}, \chi^{(3)}_{xyxx}\}$ weight terms proportional to $E_x^2 E_y$, and $\chi^{(3)}_{xyyy}$ multiplies $E_y^3$; analogously for y-polarized THG one replaces first index x with y and obtains three separate mixtures of the electric field components. It is often convenient to recast these by symmetry into $\chi^{(3)}_{xxxx}$ and $\chi^{(3)}_{yyyy}$, and the cross-coupling combinations $\chi^{(3)}_{xxyy} = \chi^{(3)}_{xyxy} = \chi^{(3)}_{xyyx}$ and $\chi^{(3)}_{yyxx} = \chi^{(3)}_{yxyx} = \chi^{(3)}_{yxxy}$, which directly control how an input polarized along x couples into a y-polarized output (and vice versa). Because the SRR geometry differentiates the gap axis from the transverse direction, $\chi^{(3)}_{xxxx}$ and $\chi^{(3)}_{yyyy}$ generally differ, and the cross-coupling sets are not the same. As the pump polarization is rotated, the relative weights of $E_x^3$, $E_y^3$, $E_x^2 E_y$, and $E_x E_y^2$ shift, leading to the multi-lobe THG patterns and the distinct angular positions and intensity contrasts between x- and y-detection observed in Fig. 4.

Taken together, Fig. 4 can be interpreted entirely within this in-plane framework: x- and y-polarized SHG intensities are governed by the quadratic combinations of $\chi^{(2)}$ with indices in {x,y}, while x- and y-polarized THG intensities are set by the cubic combinations of $\chi^{(3)}$ within {x,y}. Differences between the gap-aligned (x) and transverse (y) pathways, along with the cross-coupling terms that mix them, are responsible for the observed anisotropic polar plots and their relative rotations between detection channels.

Since the ITO spacer is modeled as isotropic for its effective nonlinear response, each physical source region (SRR surfaces, ITO interfaces, ground plate) radiates into the far field with its own dispersion and optical path, so their x- and y-polarized fields acquire different complex amplitudes and phases. Interference among these in-plane contributions is therefore expected and is directly visible in the number of lobes, their angular positions, and the x/y intensity contrast in Fig. 4. It is clear that our model that accounts for nonlinear dispersion in both metal and ITO regions in panels (e-h) reproduces the experimental trends in panels (a-d), capturing the lobe counts, their rotations within the xy plane, and the relative x versus y output levels at both pump wavelengths for SHG and THG.

To complement the full-wave approach, we use a compact analytical model of a driven nonlinear oscillator following approach in Ref. [47]. The dynamics along x and y are governed by linear inertial, damping, and restoring terms, driven by external forces along each axis, and perturbed by a weak cubic anharmonic potential. The equations of motion are given by

$$m_x \ddot{x} + \gamma_x \dot{x} + k_x x = F_x - \frac{\partial}{\partial x} U(x, y),$$

$$m_y \ddot{y} + \gamma_y \dot{y} + k_y y = F_y - \frac{\partial}{\partial y} U(x, y),$$

$$U(x, y) = a x^3 + b x^2 y + c x y^2 + d y^3, \tag{10}$$

$$E_x = E_0 \cos(\theta) e^{-i\omega t} + \text{c.c.},$$

$$E_y = E_0 \sin(\theta) e^{-i\omega t} + \text{c.c.}.$$

where the effective masses $m_i$, damping factors $\gamma_i$, and spring constants $k_i$ may differ along x and y to capture the intrinsic anisotropy of the SRR. The anharmonic correction to the potential energy $\delta U$ retains all symmetry allowed cubic terms in x and y with coefficients a, b, c, and d. Allowing $m_x \neq m_y$ is natural here: the oscillator represents a collective charge motion in an anisotropic nano-resonator rather than a single particle, so the effective inertia and damping can differ along the gap axis (x) and the transverse direction (y). While higher order terms can be included, the cubic truncation already generates the second and third harmonic responses that are the focus of Fig. 4.

We assume weak anharmonicity and construct the solution perturbatively. The leading response at the driving frequency is obtained from the linear part of (10), with the pump field parameterized by in-plane components $E_x$ and $E_y$. Substituting the linear solution into the anharmonic force generates source terms at twice and three times the drive frequency. Projecting those sources back onto the oscillator eigen-responses along x and y yields closed form expressions for the second harmonic and third harmonic source amplitudes in the xy plane. Because the detection in Fig. 4 is strictly in-plane, we keep only x- and y-polarized far-field channels.

For SHG, the in-plane source amplitudes naturally organize into quadratic angular harmonics of the pump polarization. The x-polarized and y-polarized second-harmonic intensities take the compact form

$$\left| g^{(2)}(\theta) \right|^2 = \left| A_g \cos^2(\theta) + 2 B_g \sin(2\theta) + C_g \sin^2(\theta) \right|^2, \tag{11}$$

$$g = x, y.$$

Each intensity is the squared modulus of a linear combination of the three quadratic basis functions with complex prefactors $A_i$, $B_i$, $C_i$. These prefactors include the effects of: (i) the oscillator parameters along each axis (masses, damping, and stiffness) that govern the linear near resonant response; (ii) the mapping from quadratic field products to the second harmonic sources, which reflects the in-plane $\chi^{(2)}$ tensor combinations; and (iii) the radiation and collection efficiencies into x and y polarizations. The cross-term proportional to $\sin(2\theta)$ controls the rotation and asymmetry of the two-lobe SHG patterns, while the relative magnitudes of the $\cos^2(\theta)$ and $\sin^2(\theta)$ terms set the x/y contrast and the lobe depths.

For THG, the same procedure generates cubic angular harmonics. The x-polarized and y-polarized third-harmonic intensities are

$$\left|g^{(3)}(\theta)\right|^2 = \left|D_g \cos^3(\theta) + F_g \cos^2(\theta)\sin(\theta) + G_g \cos(\theta)\sin^2(\theta) + H_g \sin^3(\theta)\right|^2, \quad (12)$$

$$g = x, y.$$

Here $D_i$, $F_i$, $G_i$, $H_i$ are complex coefficients. Physically, these coefficients encode the pure axis contributions ($\cos^3(\theta)$ and $\sin^3(\theta)$ terms), which are sensitive to differences between the gap-aligned and transverse pathways, and the remaining mixed terms that govern polarization mixing and determine the number and angular position of lobes in the THG polar plots.

A key feature of this formulation is that all fitting coefficients in (11) and (12) are treated as complex numbers. This is essential for reproducing the measured polar patterns with high fidelity. The phases of all coefficients capture several physically distinct phase delays: (i) dispersion of the nonlinear response in the contributing media (e.g., metal surfaces versus ITO interfaces) that contains different intrinsic phases to the effective $\chi^{(2)}$ and $\chi^{(3)}$ pathways; (ii) propagation and reflection phases accumulated within the multilayer stack (including the ITO spacer and the ground plate), which differ for fields launched along x and y and for different harmonic orders; and (iii) geometry dependent radiation phases associated with distinct SRR facets and current paths that weight the x and y far-field channels differently. Interference among these contributions (being constructive for some angles and destructive for others) requires nontrivial relative phases to reproduce lobe rotations, unequal lobe depths, and subtle asymmetries between x- and y-polarized outputs. Constraining the coefficients to be purely real suppresses these phase-controlled interference effects and fails to capture key experimental nuances.

In practice, we use the SHG and the THG forms in (11) and (12) to fit the experimental x- and y-polarized polar patterns in Fig. 4(a-d). To remove non-identifiability, we fix a global phase by setting the first coefficient real. The fits are carried out with variable projection: for any set of fitting coefficients, we compute the least-squares optimal background and scale overall amplitude in closed form. We then optimize the independent parameters using Levenberg-Marquardt algorithm with analytic Gauss-Newton Jacobians. Initialization is symmetry-aware (matching qualitative features of the data). We used a small multistart strategy, perturbing the initial guess with random noise to mitigate local minima. This approach reduces dimensionality, improves conditioning, and yields robust, high-quality fits while respecting the physical structure of the model. The resulting fits, shown in Fig. 4(i-l), reproduce all prominent features of the measurements: the lobe counts, their angular positions and rotations with pump wavelength, the x/y intensity contrast, and the small asymmetries between opposite lobes. This level of agreement is achieved only when the fitting parameters are allowed to be complex, thereby accommodating the relative phases among the contributing in-plane pathways. The fitted complex amplitudes are reported in Tables II (SHG) and III (THG).

Cases with complex fitting coefficients are rich in physics, as they represent a wide range of interference between fields emitted by different parts of the system. For example, consider panel (a) at a 1615 nm pump. The fitted SHG coefficients reflect simple phase relations between "same-axis" and "mixed" pathways. $A_x$ and $C_x$ (and $A_y$) are real because they are dominated by same-axis quadratic sources, $E_x \cdot E_x$ and $E_y \cdot E_y$, whose effective second order responses are nearly in phase with the respective x- and y-channel radiation at $2\omega$. By contrast, the mixed pathway $E_x \cdot E_y$ samples linear x and y oscillators that are out of phase at this wavelength (different detuning and damping) and radiates through distinct interfaces and current paths, adding propagation/reflection delays in the ITO spacer and from the ground plate. This pushes the mixed term close to quadrature (i.e., $\pi/2$ out of phase) with the same-axis terms, making $B_x$ nearly purely imaginary. In the y channel, mixed coupling is weaker, so $B_y$ is small and complex, while the dominant $E_y$-driven contribution projects onto y with an additional phase delay set by the y-channel radiation conditions, rendering $C_y$ nearly purely imaginary. This real/imaginary pattern is the compact signature of the underlying relative phases that the fit exploits to reproduce the measured lobe orientations, depths, and x/y contrast.

**Table II**

|         | $A_x$  | $B_x$              | $C_x$             | $A_y$  | $B_y$              | $C_y$                |
|---------|--------|--------------------|-------------------|--------|--------------------|----------------------|
| 1615 nm | 0.356  | -0.0207+i·0.922    | 0.145+i·0.0281    | 0.964  | 0.0542-i·0.0472    | -0.00939+i·0.256     |
| 1300 nm | 0.0869 | -0.456+i·0.866     | 0.136+i·0.128     | 0.744  | 0.156-i·0.00975    | 0.563-i·0.326        |

Fitting parameters for (11) corresponding to panels (i) and (j) in Fig. 4.

The complex x-polarized SH polarization (panels (a), (e), and (i) in Fig. 4, black lines) associated with the cross term adds in quadrature to the direct x→x and y→x projections. Physically this phase offset arises because the mixed source samples, which carry different driven phases due to detuning and damping, and then radiating through a stack that imparts additional Fresnel and propagation phase at $2\omega$. The combination of cross-oscillator mixing and multilayer phase typically drives the $\sin(2\theta)$ coefficient toward $\pm i$ relative to the same-axis terms.

**Table III**

|         | $D_x$  | $F_x$           | $G_x$          | $H_x$          | $D_y$   | $F_y$           | $G_y$           | $H_y$          |
|---------|--------|-----------------|----------------|----------------|---------|-----------------|-----------------|----------------|
| 1615 nm | 0.71   | -0.027+i·0.68   | -0.20+i·0.0086 | 0.029-i·0.019  | 0.019   | 0.81-i·0.010    | 0.018-i·0.48    | -0.33-i·0.046  |
| 1300 nm | 0.52   | -0.020+i·0.47   | -0.69-i·0.11   | 0.024+i·0.13   | -0.0053 | -0.60+i·0.091   | 0.056-i·0.0082  | 0.78-i·0.12    |

Fitting parameters for (12) corresponding to panels (k) and (l) in Fig. 4.

Additionally, this quadrature explains the "butterfly" shape of panel (i). If $B_x$ were real and comparable in magnitude to $A_x$ and $C_x$, interference along certain angles would be maximally constructive or destructive, producing deep nodes on the 0-180° cut. When $B_x$ is imaginary, the mixed term does not directly cancel the same-axis terms; instead, it rotates the vector sum in the complex plane. The polar lobes remain two-fold symmetric, but the nominal node along 0-180° lifts because the imaginary mixed contribution cannot destructively interfere with the real background. Any small real leakage from the same-axis x→x SHG or slight amplitude/phase imbalance between ±$k_x$ components further prevents exact cancellation, yielding the finite intensity seen along that axis. The difference between panel (i) and the numerical panel (e) follows naturally from this picture. The numerical model enforces ideal symmetries and simpler 2ω phases, effectively making $B_x$ real or 0 relative to the chosen radiation reference, so destructive interference along 0-180° was exact and the curve crossed zero. In the measurement or fitted case, realistic dispersion and stack-induced phases add a small background and rotate the mixed pathway into quadrature, lifting the node while preserving the overall symmetry.

Another representative example is Fig. 4, panel (d) showing THG response at a 1300 nm pump. The black curve's multi-lobe structure is captured with $D_x$ real, $F_x$ nearly purely imaginary, and $G_x$, $H_x$ fully complex because the underlying cubic pathways acquire different relative phases. $D_x$ corresponds to the pure $E_x \cdot E_x \cdot E_x$ route into x-polarized 3ω and is phase-aligned with the x-channel radiation at 3ω, so it appears real. $F_x$ weights $E_x \cdot E_x \cdot E_y$; mixing the near-resonant x response with a detuned y response and routing through distinct current paths adds roughly a quarter-cycle phase via anisotropic detuning and spacer/ground plate reflections, pushing $F_x$ close to quadrature and thus nearly imaginary. $G_x$ ($E_x \cdot E_y \cdot E_y$) and $H_x$ ($E_y \cdot E_y \cdot E_y$ projected into x) sample two y-axis responses and project through geometry-dependent polarization mixing; they track different 3ω radiation and Fresnel pathways than $E_x \cdot E_x \cdot E_x$ and $E_x \cdot E_x \cdot E_y$, accumulating independent propagation and reflection phases. As a result, neither is locked to 0 or 90 degrees, and both emerge fully complex. These phase relations are what let the fit in panel (l) reproduce the detailed lobe positions, depth asymmetries, and envelope of the THG pattern. We note that our numerical model (panel (h)) nicely replicates this behavior as well.

**Conclusion**

We investigated a metasurface comprising a gold ground plate, an ITO spacer with an ENZ wavelength at 1615 nm, and a periodic array of split-ring resonators. The study combined nanofabrication with linear and nonlinear optical characterization in reflection, and a multi-tiered modeling approach. Metal was described by a hydrodynamic electron model self-consistently coupled to our in-house FDTD solver. For ITO, we progressively increased model complexity: (1) a linear dispersive dielectric; (2) adding static $\chi^{(2)}$ and $\chi^{(3)}$; and (3) adding dispersive $\chi^{(2)}$ and $\chi^{(3)}$ represented by Lorentzian profiles. The full dispersive nonlinearity in (3) reproduces the measurements with high fidelity across both second- and third-harmonic responses, including angularly resolved patterns and absolute trends.

A key outcome is that dispersive nonlinearities in ITO are essential. Models with static $\chi^{(2)}$ and $\chi^{(3)}$ fail to capture the measured SHG/THG amplitudes, lobe orientations, and node depths, whereas introducing Lorentzian dispersion in $\chi^{(2)}(\omega;2\omega)$ and $\chi^{(3)}(\omega;3\omega)$ yields the correct complex phase relationships between mixed and same-axis pathways. This dispersion-driven phase rotation explains lifted nodes and lobe rotations in the angular scans and is necessary to reconcile experiment and theory near the ENZ regime, where small spectral shifts strongly modulate both magnitude and phase of the nonlinear response.

Additionally, we developed a driven nonlinear oscillator model that yields closed form angular dependences for SHG and THG. Fitting the measured polar scans with complex coefficients, the analytical model captures the observed lobe rotations, node lifting, and contrast. The complex nature of the fitting parameters naturally arises from phase differences between field components launched and re-radiated by distinct pathways (metal, ITO bulk, and interfaces) subject to dispersion and propagation at the fundamental and harmonic frequencies. Consistency between the numerical hydrodynamic-FDTD simulations and the analytical fits confirms this phase-interference picture and explains why mixed pathways can be effectively in quadrature with same-axis contributions.

Looking forward, the analytical framework offers a practical inverse-design tool. By targeting desired polar signatures or harmonic yields, one can solve for the complex coefficients and back-out structural or material adjustments, such as ITO thickness, resonance detuning, or dispersion engineering, that realize those coefficients. Coupled with rapid numerical validation, this enables efficient optimization of ENZ-assisted metasurfaces for tailored nonlinear beam patterns, enhanced conversion efficiency, and functional angular responses, extending to multiplexed sources and phase-encoded harmonic wavefronts.

**Acknowledgements**

M.S. is grateful to the U.S. Department of the Air Force Summer Faculty Fellowship Program for funding during summers of 2024 and 2025. M.S. research is funded by the Air Force Office of Scientific Research under Grants No. FA9550-22-1-0175 and No. FA9550-25-1-0096. Development of numerical FDTD tools is sponsored by the Office of Naval Research, Grant No. N000142512090. All simulations were made possible through the DOD High Performance Computing Modernization Program and were performed on DOD HPC systems: Narwhal, Nautilus, and Blueback. S.V. and J.R.H. acknowledge support from the Air Force Office of Scientific Research under Grants No. FA9550-23RYCOR001 and No. FA9550-25RYCOR006. The research performed by C.E.S. and M.K. at the Air Force Research Laboratory was supported by contract award No. FA807518D0015.